\documentclass{article}
\usepackage[dvips]{graphicx}

\title{Determination of the accuracy of measuring the energy of charged
 particles by their path lengths in nuclear emulsion}

\author{A.S.~Barabash$^1$, V.Ya.~Bradnova$^2$, M.M.~Chernyavsky$^3$, \\ 
V.V.~Dubinina$^1$, N.P.~Egorenkova$^1$, S.I.~Konovalov$^1$, \\ 
N.G.~Polukhina$^3$, E.A.~Pozharova$^1$, T.V.~Shchedrina$^3$, \\
V.A.~Smirnitsky$^1$, N.I.~Starkov$^3$, Tan Nang So$^3$, \\
 and V.I.~Umatov$^1$\\ [0.4cm]
$^1${\it\small Institute of Theoretical and Experimental Physics,}\\
{\it\small B.~Cheremushkinskaya 25, 117218 Moscow, Russia} \\
$^2${\it\small Joint Institute for Nuclear Research, 141980 Dubna, Russia} \\
$^3${\it\small Lebedev Physical Institute, Russian Academy of Sciences,}\\
{\it\small Leninsky Prospekt 53, 119991 Moscow, Russia}}

\date{ }

\begin{document}

\maketitle

\begin{abstract}

The path lengths of monochromatic muons emerging in the $\pi \to \mu\nu_{\mu} $ decay
were measured by the coordinate method with the view to determine their energy by 
path length. The dispersion of muon energy measurements by this method was 
$ \sigma_{\mu} = (0.11 \pm 0.01)$ MeV, which corresponds to the accuracy of 
$ \sigma \approx 2.7\%$ of the energy of a charged particle estimated by its path length in 
nuclear emulsion. The developed method will enable measurements of electron energies 
in $2\beta $ decay ($\sim 3$ MeV) to an accuracy of $ \sigma \approx 5\%$.

\end{abstract}



The feasibility of using nuclear emulsions with molybdenum filling to search for 
$\beta\beta$ decay was shown in \cite{ASH10}. Assessment of the background conditions 
done in \cite{ASH11} showed the possibility of achieving  
the sensitivity to $0\nu\beta\beta$ decay of $^{100}$Mo at the level of 
$\sim 1.5 \cdot 10^{24}$ years in one year of measurements for 1 kg of $^{100}$Mo.

An important factor in the success of this work is the ability of 
automatic processing of emulsions, i. e. the search for and recognition of tracks 
of particles at the PAVICOM facility. A major problem in the recognition of tracks 
of particles is  that of their separation from the background. One of the background 
processes is the decay of radioactive nuclei occurring in the emulsion with the 
simultaneous emission of electrons. The main feature of such reactions is the presence
 of characteristic spatial configurations - stars (Fig. 1a). Their pre-selection is
carried out on the basis of a characteristic feature, a small value of the ratio of 
the number of pixels that form a star to the area of the circumscribed rectangle 
(Fig. 1b).  This procedure is performed at the stage of clustering (singling-out 
of darkening areas of a given level on a digitized image). Further on, to recognize 
the inner structure of a star (its rays, their positions on the image, etc.), a grid 
of mutually perpendicular lines is superimposed on the image. Points of intersection 
of the star and grid lines form the skeleton of the star, i.e. the set of segments 
delineating the region of its positioning (Fig. 2a). 

\begin{figure}
\begin{center}
\includegraphics[height=4cm]{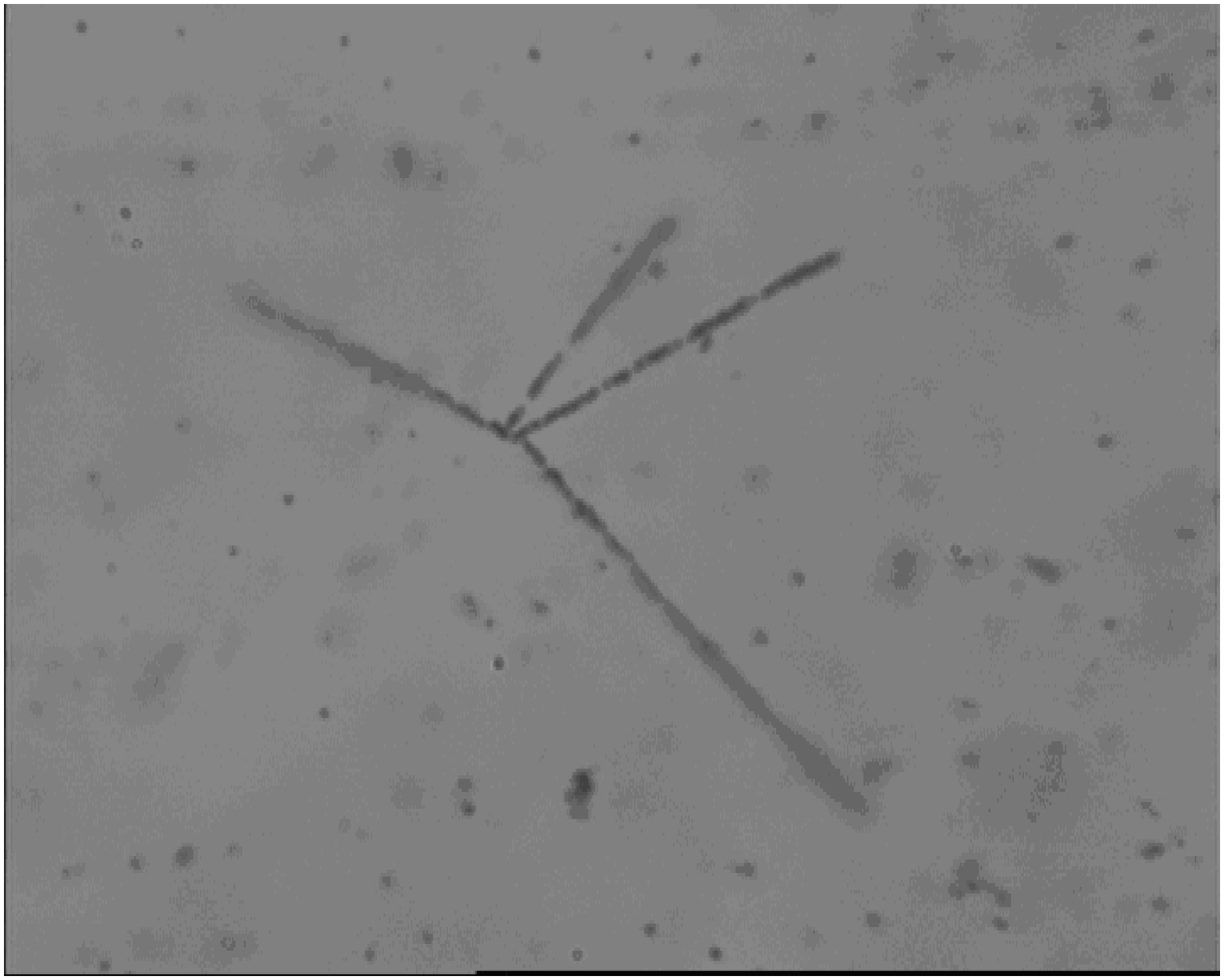} 
\hspace{5mm}
\includegraphics[height=4cm]{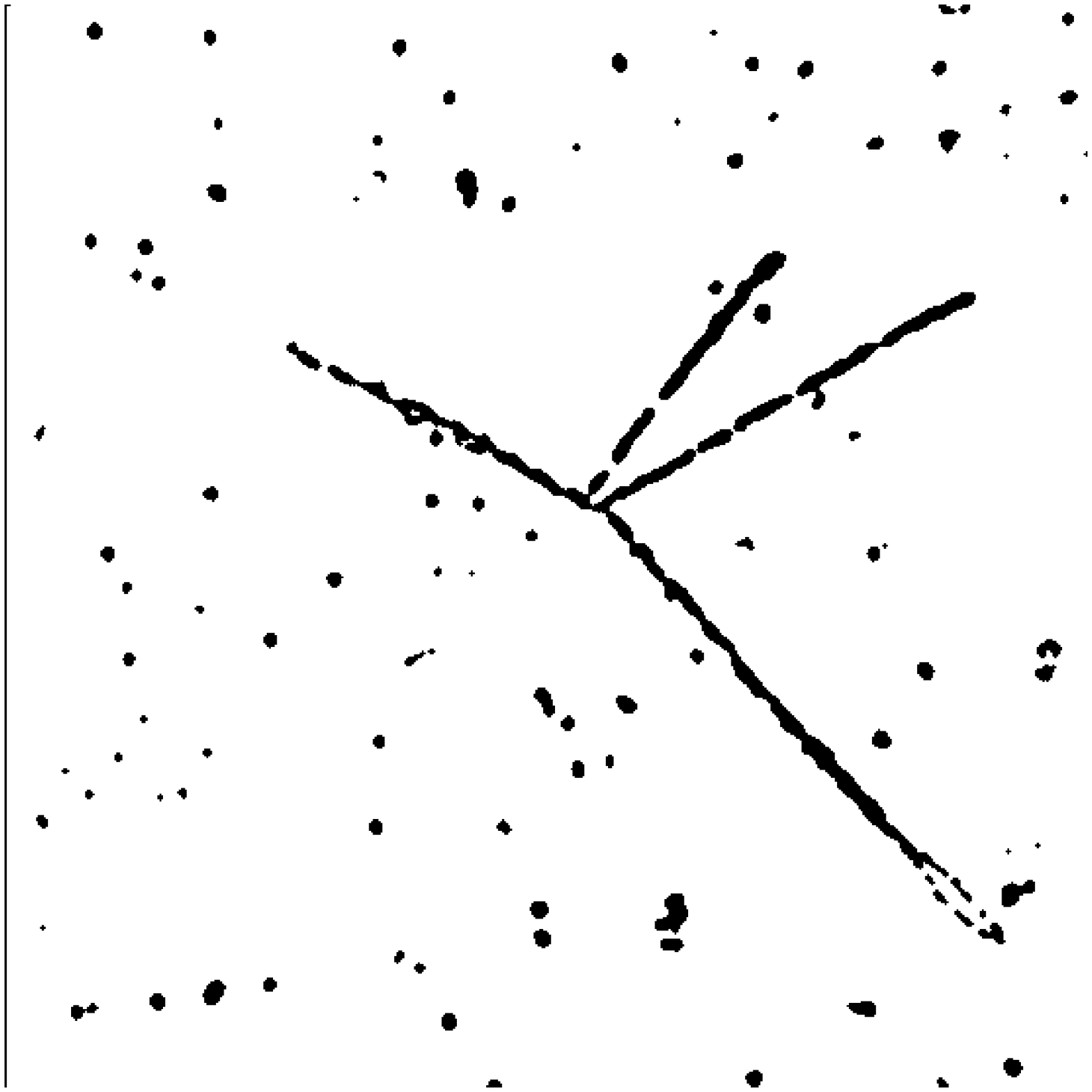}
\caption{(a) The initial image of a star. (b) The result of clustering.}
\end{center}
\label{fig:fig1}
\end{figure}

\begin{figure}
\begin{center}
\includegraphics[height=3cm]{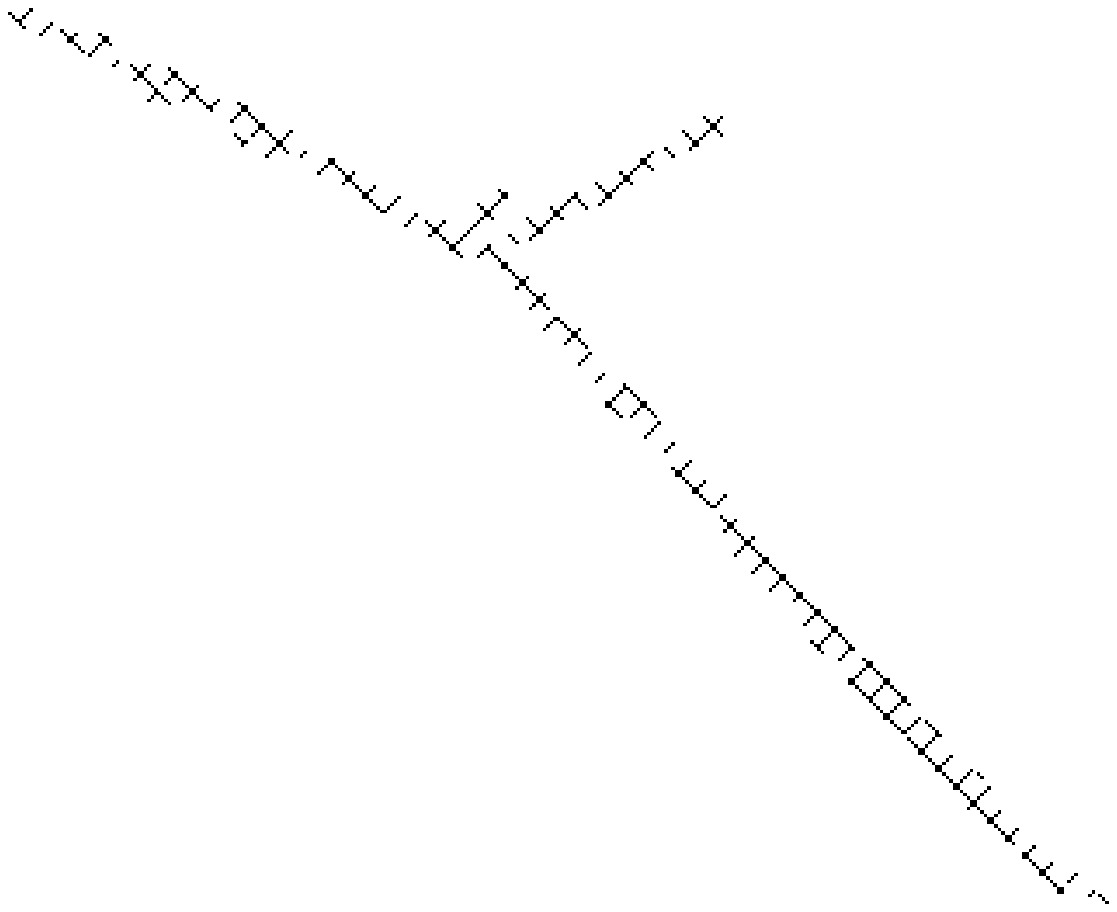} 
\hspace{5mm}
\includegraphics[height=4cm]{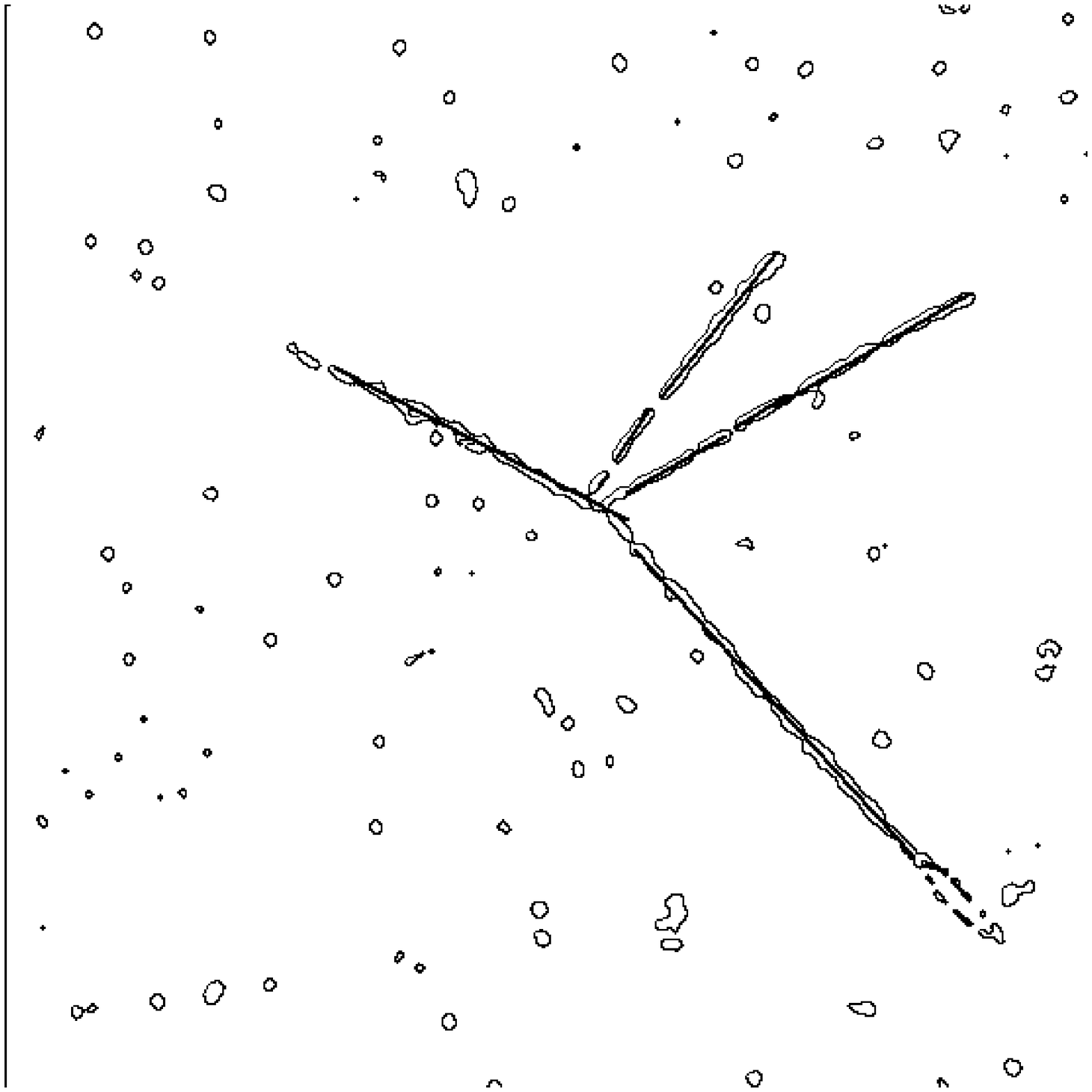}
\caption{(a) An auxiliary grid on the image of a star. (b) A result of the search 
for the axes of star rays.}
\end{center}
\label{fig:fig2}
\end{figure}

 To search for the positions of the rays closely positioned segments are grouped and 
straight lines – axes of rays – are drawn through their centres of masses. The region
 of mutual intersection of the axes corresponds to the position of the vertex of a decay.
 The positions of the rays are refined with account for the knowledge of the vertex. 
Because of non-homogeneous darkening of various parts of a star on the image, 
during the processing some rays break down into separate parts, which should be 
re-assembled for the total reconstruction of the decay star. For this 
the characteristics of clusters positioned near the star are to be analyzed. 
If the axis of a cluster coincides with the axis of a star ray the cluster is 
considered to be the continuation of the ray. As the result, the complete geometry 
of a star is formed. 
Figure 2b shows the result of applying this algorithm to the image of Figure 1a. 

The decays of radioactive nuclei are a background process and the aim of recognizing a star
 is to reject such events. However, this algorithm can also be used if a comprehensive 
analysis of events with decays of nuclei is required. Thus, to reconstruct the spatial 
geometry of a star, it is necessary to scan images in higher and lower layers and to 
search for continuations of rays on them. The authors continue to develop such
 a specialized software. 

The potential of this experiment also depends on the accuracy of measuring the energy 
of electrons which is determined from their path lengths. For calibration monochromatic 
electrons of energy $\sim 1$ MeV from a radioactive source of $^{207}$Bi are usually used.  

At present, domestic nuclear emulsions are at the stage of technology recovery by Slavich 
Joint Stock Company and can not yet be used for experiments. For this reason, to determine 
the accuracy and to develop the method of measuring the energy of charged particles 
from their path lengths, we used NIKFI BR-2 emulsion irradiated in 1967 by slow pions 
at the JINR accelerator. At the stop, the $\pi^+$ meson decays according to the pattern 
observed with 100\% efficiency in the emulsion

\begin{center}
$\pi^+ \to \mu^+ + \nu_{\mu} ; \hspace{5mm}  \mu^+ \to e^+ + \nu_{e} + \nu_{\mu} $ 
\end{center}

This decay brings about strictly monochromatic muons of energy of 4.12 MeV.  The path 
length of these muons in the standard BR-2 emulsion is 600 $\mu$m and depends on a range 
of causes: fluctuations of energy losses (due to various reasons), straggling, moisture 
emulsion  content and, which is probably the main cause, manufacturing quality, which has 
an effect on the unevenness of emulsion density along the trajectory of a charged particle. 
The latter cause does not in practice yield to control. The value of muon path length can be
 affected by decays-on-the-fly. To assess their effect, it is necessary to compare the 
lifetimes of pions and muons with their stopping times to standstill. For pions, 
the stopping time in the field of vision of a microscope (120 $\mu$m) 
is $0.0034 \cdot 10^{-9}$ s [3], and the lifetime $\tau_{\pi} = 2.6 \cdot 10^{-8}$ s [4], 
which is by $\sim 4$ orders of magnitude greater. The analogous values for muons 
(path length, 600 $\mu$μm) are $0.0095 \cdot  10^{-9}$ s and 
$\tau_{\mu} = 2.2 \cdot 10^{-6}$ s. At that ratio of stopping times and lifetimes one can 
expect less than one decay of $\pi$ and $\mu$ in flight in our statistics.

A stopped pion decays isotropically, so muons flying out at various angles are measured,
 and the only criterion of selection is the condition that the entire path length of a muon 
be in one emulsion layer. Figure 3 shows micrographs of three $\pi\mu$e decays positioned 
in the focal plane of the objective lens (field of vision, $\sim 120 \mu$m). This made it 
possible to arrange fragments of micrographs 
of one event in one micrograph (we took material from the atlas published in the appendix to [5]).  

\begin{figure}
\begin{center}
\includegraphics[height=15cm]{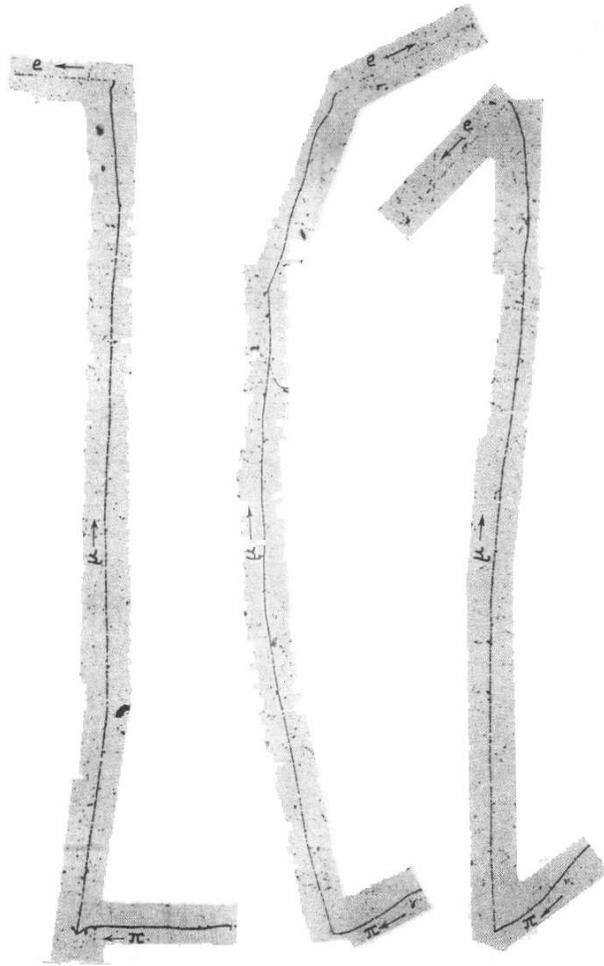} 
\caption{Micrographs of three events of a $\pi\mu$e decay in the focal plane of the objective lens 
(from the atlas published in [5]).}
\end{center}
\label{fig:fig3}
\end{figure}

Prior to its stopping, a 4-MeV muon is noticeably scattered, especially by the end of the path, 
so for the path length to be accurately measured one should carefully track changes in the 
trajectory of the muon. Measurements were done by the coordinate method. 
Depending on the geometry of the trajectory and on the scattering, the $x_i, y_i, z_i$ 
 coordinates are measured on the track of a muon for 40--100 points and the length of the 
muon track is calculated as the sum of segments between these points. Figure 4 presents 
a distribution of the number of measurement points, n, on the track of a muon ($<n> = 65$). 
This technique enabled us to improve the accuracy of determining the energy of charged 
particles from their path lengths. The measurements were carried out on a KSM microscope 
linked to a computer. Computer software made possible the operative control of the measuring results. 

\begin{figure}
\begin{center}
\includegraphics[height=7cm]{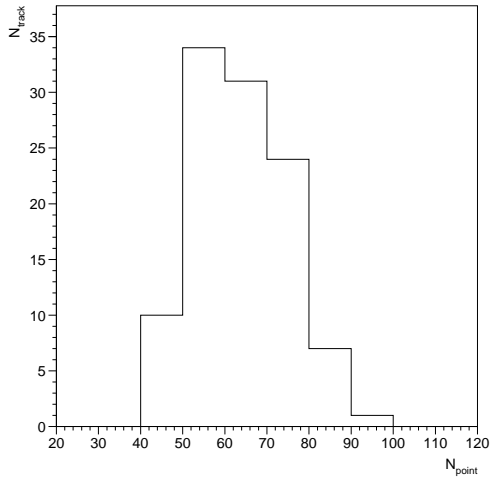} 
\caption{Distribution of coordinate measurement points on muon tracks.}
\end{center}
\label{fig:fig4}
\end{figure}

\begin{figure}
\begin{center}
\includegraphics[height=5cm]{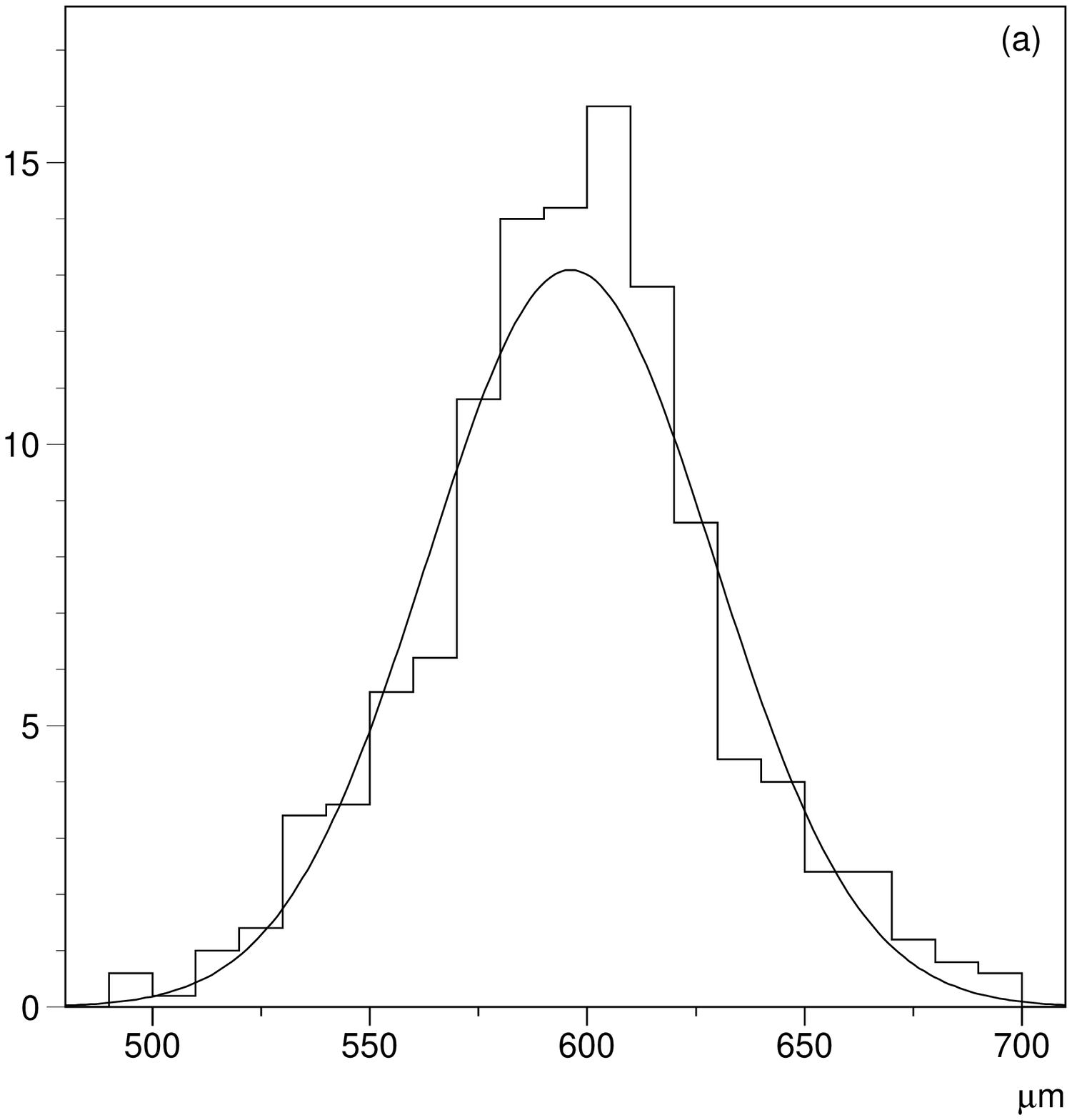} 
\hspace{5mm}
\includegraphics[height=5cm]{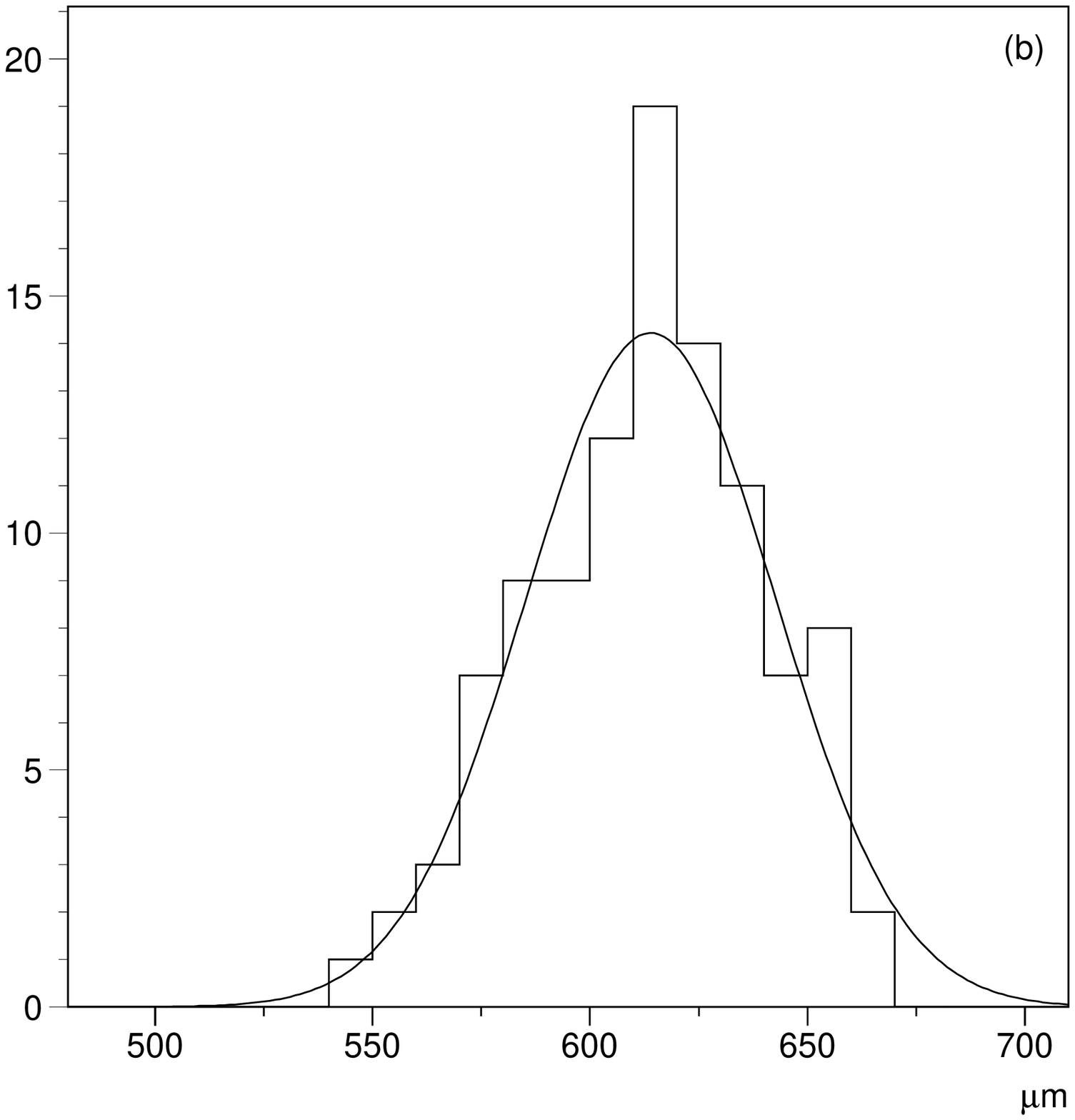}
\caption{ (a) A path length histogram for muons emerging from the  $\pi^+ \to \mu^+ + \nu_{\mu}$
 decay in [5]; (b) the same for our measurements.}
\end{center}
\label{fig:fig5}
\end{figure}

Figure 5a shows the results of the best (in our opinion) measurements of path lengths for muons  emerging in the $\pi^+ \to \mu^+ + \nu_{\mu}$ [5] and Figure 5b presents the results of our measurements. Both histograms are described by the Gaussian distribution with the following parameters: 

\begin{table}[h]
\begin{tabular}{lll}
The histogram from [5]: & $<R_{\mu}> = (596.2 \pm 1.0) ~\mu$m & $\sigma_R = (33.0 \pm 1.0) ~\mu$m \\
Our measurements:       & $<R_{\mu}> = (614.0 \pm 1.0) ~\mu$m & $\sigma_R = (28.7 \pm 0.9) ~\mu$m            
\end{tabular}
\end{table}

The dispersion of our measurements is smaller than that obtained in [5].

Calculations to convert  path lengths of various charged particles in nuclear emulsion to 
their energies for Ilford G-5 emulsion was published in [5]. Manufacturers of nuclear 
emulsions preserve the nuclear compositions constant, so theoretically the path length--energy 
ratios should be constant for a given type of emulsion. As the experience shows [5], at a change of the moisture content from 20 to 60\% 
the path length--energy calibration measurements remain practically invariable. Within the range of energies from 1 to 50 MeV the path length--energy dependence for charged particles in Ilford G-5 emulsions was found and verified to be as follows: 

\begin{eqnarray*}
E_t &=& a_t (R_t)^k \\             
E_d &=& a_d (R_d)^k \\
E_p &=& a_p (R_p)^k \\         
E_K &=& a_K (R_K)^k \\
E_{\pi} &=& a_{\pi} (R_{\pi})^k \\
E_{\mu} &=& a_{\mu} (R_{\mu})^k
\end{eqnarray*}  
       
Here, R is measured in $\mu$m, E in MeV, the index of power k is the same for all particles, k = 0.581; the indices at energies, path lengths and multipliers correspond to particles: t -- tritium, d -- deuterium,
 p -- proton, K -- kaon, $\pi$ -- pion; $\mu$ -- muon. The measured values are [5]: $a_t = 0.398$, 
$a_d = 0.336$, $a_p = 0.251$, $a_K = 0.192$, $a_{\pi} = 0.113$, $a_{\mu} = 0.101$.

Insignificant changes in the nuclear composition of emulsions affect the values of coefficient $a_i$. For this reason it is required to calibrate emulsions by irradiating them with known particles of certain energy. In our measurements, by the results of calibration the coefficient for BR-2 emulsions was found to be 
$a_{\mu} = 0.0989$.

\begin{figure}
\begin{center}
\includegraphics[height=7cm]{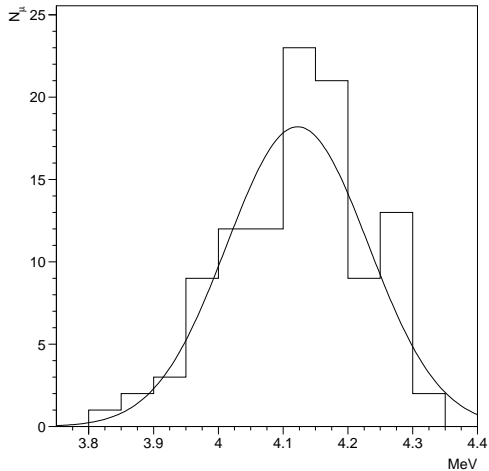} 
\caption{Muon energy distribution obtained from calculations by relation (1) with coefficient 
$a_{\mu} = 0.0989$.}
\end{center}
\label{fig:fig6}
\end{figure}

Figure 6 shows an energy distribution of muons emerging at the decay of stopped pions. The distribution was calculated corresponding formula above. The result of fitting the histogram by the Gaussian distribution yields the values: 

$$   <E_{\mu}> = (4.12 \pm 0.01) ~{\rm MeV}  \hspace{1cm}   \sigma_{\mu} = (0.11 \pm 0.01)~{\rm MeV} $$

The non-symmetric distribution with respect to mean energy is due to the fact that 
the path length determined as the sum of rectilineal segments is always smaller than 
the true length of a trajectory. The calculation of a root mean square error from 
the results of Fig. 6 yields the value  $\sigma_{\mu} = (0.104 \pm 0.014)$  MeV, which is 
close to that obtained by the fitting.

Thus, we have determined the accuracy of measuring the energy of a charged particle 
(muon) from its path length in nuclear emulsion. This result is more accurate than 
the result of path length--energy measurement for muons, $\sigma_{\mu} \approx 4\%$, 
made in [5]. However, the experiment is to measure two electrons of total energy up 
to 3  MeV. Normal emulsions have a sensitivity of $\sim (30-50)$ grains per 100 $\mu$m that 
makes it possible to trace sufficiently accurately the trajectory configuration of an 
electron along its entire path till standstill. The loss of accuracy at this stage of 
measurements can not be significant. The greatest problems will occur in passing from 
layer to layer of the emulsion chamber. This stage is inevitable as the path length of 
electrons for the $\beta\beta$ decay of $^{100}$Mo can reach 5 mm (at $E_e = 3$ MeV), 
and the thickness of the emulsion layers in the chamber is $\sim 500 \mu$m. 
In passing from layer to layer, the last observed grain of a track will practically 
always be not on the surface but somewhere inside the emulsion. Therefore, in each 
passage the trajectory will be extrapolated two times to determine the points of 
intersection of two surfaces by the electron. The radiation correction to electron 
energy is $\sim 1\%$, as this energy is much less than the critical value. Emulsion 
density fluctuations along the trajectory will affect the path length much less 
due to a better averaging of inhomogeneities at a trajectory length of several
 millimeters.  All these and, probably, other sources increasing the error of measuring 
the energies of electrons from their path lengths can be reliably found by irradiating 
small emulsion chambers with electrons of exactly known energy. According to our 
preliminary estimates, the accuracy of determining the energy of electrons from their 
path lengths can be $\sigma \sim 5\%$.

The work was partially supported by the RFBR grant No 11-02-00476.

 --------------------------------------------------------------

\end{document}